\documentclass[pop,twocolumn, groupedaddress, nofootinbib]{revtex4}
\usepackage{amsmath}
\usepackage{epsfig}
\usepackage{epstopdf}
\usepackage{graphicx}          
\graphicspath{{figures/}}
\usepackage{caption}

\usepackage{graphicx}
\usepackage{epsfig}
\usepackage{times}
\usepackage{amsmath}
\usepackage{cancel}
\usepackage{pifont}

\def\psit{\tilde{\psi}}

\def\simleq{\mathrel{\smash{\mathop{\raise2pt\hbox{$<$}}\limits_%
   {\smash{\raise4pt\hbox{$\sim$}}}}\vphantom\leq}}
\def\simgeq{\mathrel{\smash{\mathop{\raise2pt\hbox{$>$}}\limits_%
   {\smash{\raise4pt\hbox{$\sim$}}}}\vphantom\geq}}

\def\psit{\tilde{\psi}}

\def\psit{\tilde{\psi}}

\def\simleq{\mathrel{\smash{\mathop{\raise2pt\hbox{$<$}}\limits_%
   {\smash{\raise4pt\hbox{$\sim$}}}}\vphantom\leq}}
\def\simgeq{\mathrel{\smash{\mathop{\raise2pt\hbox{$>$}}\limits_%
   {\smash{\raise4pt\hbox{$\sim$}}}}\vphantom\geq}}

\renewcommand{\vec}[1]{\ensuremath{\mathbf{#1}}}

\newcommand{\comment}[1]{}
\renewcommand{\baselinestretch}{1}

\def\simleq{\mathrel{\smash{\mathop{\raise2pt\hbox{$<$}}\limits_%
   {\smash{\raise4pt\hbox{$\sim$}}}}\vphantom\leq}}
\def\simgeq{\mathrel{\smash{\mathop{\raise2pt\hbox{$>$}}\limits_%
   {\smash{\raise4pt\hbox{$\sim$}}}}\vphantom\geq}}

\renewcommand{\subsectionmark}[1]{\markright{\S~\thesubsection}}
\renewcommand{\sectionmark}[1]{\markright{\S~\thesection}}
\renewcommand{\baselinestretch}{1}

\begin{document}

\title{The scaling of forced collisionless reconnection}
\author{Brian P. Sullivan}
\email{brian.sullivan@unh.edu}
\author{Barrett N. Rogers}
\affiliation{Department of Physics and Astronomy, 
Dartmouth, Hanover, NH 03755}
\date{\today}

\begin{abstract}
This paper presents two-fluid simulations of forced magnetic reconnection with
finite electron inertia in a collisionless three-dimensional (3D) cube with
periodic boundaries in all three directions. Comparisons are made to
analogous two-dimensional simulations. Reconnection in this system is
driven by a spatially localized forcing function that is added to the
ion momentum equation inside the computational domain.  Consistent
with previous results in similar, but larger forced 2D simulations
[B. Sullivan,   B. N. Rogers,  and   M. A. Shay,  Phys. Plasmas \textbf{12}, 122312 (2005)], for sufficiently strong forcing the reconnection
process is found to become Alfv\'enic in both 2D and 3D, $i.e.$, the
inflow velocity scales roughly like some fraction of the Alfv\'en
speed based on the upstream reconnecting magnetic field, and the
system takes on a stable configuration with a dissipation region
aspect ratio on the order of $0.15$.
\end{abstract}


\maketitle

\section{Introduction}
Magnetic reconnection allows a magnetized plasma system to convert
magnetic energy into high speed flows and thermal energy, and is the
main driving mechanism behind energetic phenomena such as solar
flares, magnetospheric substorms, and tokamak sawteeth. A defining
feature of these phenomena is that they are very bursty: in a
substorm, for example, a significant fraction of the lobe magnetic
flux is reconnected over a period of about 10 minutes, while the time
between periodic substorms is roughly 2.75 hours
\cite{borovsky}. Similarly, in the case of an X-class solar flare, the
energy release may occur during several minutes or less in active
regions that can last for weeks before reconnecting \cite{miller}.  A
viable model of reconnection in such systems must therefore be
consistent with the relatively long, quasi-stable periods in between
reconnection bursts, as well as the rapidity of the energy release
once a burst has somehow been triggered.

Regarding the rapidity of reconnection, recent research (see, $e.g.$, Refs.
\cite{Birn01, BirnHesse01, Ma01, Bhattacharjee01, Pritchett01, Rogers01, Shay4, 
Zeiler02, Scholer03, Huba04, Fitzpatrick04, Ricci04, Shay04, Cassak, Chacon07})
has suggested that the reconnection process can indeed generate 
sufficiently fast rates of energy release to be consistent 
with observations provided that various conditions are met.  
For example, studies of spontaneous, unforced reconnection 
in systems with sufficiently narrow current sheets ($\delta \simleq d_i= c/\omega_{pi}$ in the case without a guide magnetic field) have found that the reconnection process becomes
Alfv\'enic - that is, the inflow velocity of plasma and magnetic flux
into the reconnection region scales roughly like some fraction of the
Alfv\'en speed based on the upstream reconnecting component of the
magnetic field. The narrowing of the current layer down to widths
$\delta \simleq d_i$ serves as a trigger point for the onset of fast
reconnection in this case due to the important role played by
non-ideal effects such as the Hall term \cite{Birn01}.  Without the
addition of ad-hoc anomalous resistivity terms, simulations of systems
with substantially thicker current sheets do not exhibit such fast
reconnection behavior: at best the reconnection rates are throttled by
the formation of long, narrow outflow layers as in the Sweet-Parker
model, or in unfavorable magnetic geometries ($e.g.$ $\Delta'\leq 0$
~\cite{FKR}) no spontaneous reconnection may take place at all.

The fact that reconnection can occur very rapidly or not at all is
consistent with the burstyness of the reconnection phenomena discussed
earlier provided one can explain how a system can transition from a
quasi-stable phase (characterized, for example, by a relatively thick
current sheet) to a period of fast reconnection (a narrow current
sheet).  One possible mechanism to explain this transition is explored
in this paper: to externally force a stable, thick current sheet
system so that the magnetic flux and electric current are compressed in a
localized region down to the small scales $\delta \simleq d_i$ at
which non-ideal effects such as the Hall term become important. This
compression is generated in our simulations by a spatially localized
external forcing term that is added to the ion momentum equation
inside a three-dimensional (3D) computational domain.  This method allows the scaling of the reconnection rate in 3D to be studied over a wide dynamic range:
by changing the level of forcing, our simulations achieve more than an
order of magnitude variation in the magnetic field just upstream of
the reconnection region, and more than two orders of magnitude
variation in the reconnection rate. The simulations show that such
localized forcing in 3D can indeed induce fast rates of reconnection
similar to those observed in 2D, spontaneously reconnecting, narrow
current sheet systems, even in magnetic geometries that would not reconnect at all in the absence of forcing. This result - that a system with
a locally narrow current sheet will exhibit fast reconnection behavior
(whether the narrow current sheet is created by external forcing or is
present initially) - is reasonable but not obvious, since the magnetic
geometries outside the forced zone in the two cases may be very
different.

This finding extends to 3D a similar result that was obtained in a
larger 2D study \cite{Sullivan}, as well as an earlier 2D multi-code
study of forced reconnection known as the Newton Challenge
\cite{NEWTON, Pritchett05, Huba06}.  The simulations in the Newton Challenge were
initialized with a relatively narrow current sheet that was unstable
to spontaneous reconnection but was somewhat thicker than the
threshold required for fast (Alfv\'enic) reconnection (two ion skin
depths compared to roughly one). The current sheet was then pinched by
a spatially and temporally dependent inflow velocity imposed at the
upstream boundary of the simulation domain.  The result was that
Alfv\'enic reconnection indeed occurred in all studies that included
the Hall term, albeit with reconnection rates $\sim 20\%-50\% $ lower
than those found in studies of spontaneous (unforced) reconnection in
systems with somewhat narrower current sheets (one ion skin depth).
The work described here goes beyond the Newton Challenge to explore,
over a wide range of parameters, the scaling of forced magnetic
reconnection in a tearing mode stable ($\Delta' \le 0$) system with an
initially broad current sheet and a forcing function that is spatially
localized in 3D.

Several limitations of the present work that are worth noting. 
Our simulations explore the dynamics of forced reconnection in
a cubical box of 25.6 ion skin depths (the 3D analog of the GEM
Challenge system \cite{Birn01}) using a simple isothermal two-fluid model \cite{Shay4}
and periodic boundary conditions. The frozen-in
condition is broken at the smallest scales in this model by electron
inertia and numerical diffusion effects. Particle simulations of
reconnection, on the other hand, have shown that non-gyrotropic
pressure tensor effects, which are not included here, should play the
dominant role in this regard. Past 2D studies of unforced reconnection
(see, for example, Refs.~\cite{Birn01, BirnHesse01, Ma01, Bhattacharjee01, Pritchett01, Rogers01}) 
have suggested that the reconnection rate is
not strongly sensitive to this difference, and it is hoped that this
insensitivity carries over to the present study. In addition, 2D
particle simulations in much larger boxes are currently being used to
study the sensitivity of the reconnection process to the boundary
conditions and the physics of the electron diffusion region
\cite{Bessho05, Karimabadi07, Shay07, Drake08, Pritchett08}. 
Due to the size limitations imposed by the 3D nature of our
study, as well as the physical simplicity of our model, these
important issues lie outside the scope of the present work.

The organization of this paper is as follows. In section \ref{model}
we describe the simulation model, equilibrium profiles, and forcing
function.  In section \ref{results} we present and discuss the results
of the simulations included in this study. The main conclusions are
summarized in section \ref{conclusions}.

\section{Simulation Model}
\label{model}
Our simulations are based on the two-fluid equations with the addition
of a spatially localized forcing term in the ion momentum equation. These
equations in normalized form are \cite{Shay04}:

\begin{equation}
n\left(\partial_t\vec{V}_i+\vec{V}_i\cdot\nabla\vec{V}_i\right)
=\vec{J}\times\vec{B}-\nabla p+F_y\hat{\vec{y}} 
\label{mom}
\end{equation}

\begin{equation}
\partial_t\vec{B}'=-\nabla\times \vec{E}'
\label{faraday}
\end{equation}

\begin{equation}
\vec{E}'+\vec{V}_i\times\vec{B}=\frac{1}{n}
\left(\vec{J}\times\vec{B}'-\nabla p_e \right)
\label{ohm}
\end{equation}

\begin{equation}
\partial_t n +\vec{V}_i\cdot\nabla n=-n\nabla\cdot \vec{V}_i
\label{den}
\end{equation}

\begin{equation}
p_e=nT_e\ ,\quad p_i=nT_i\ ,\quad p=p_i+p_e
\label{pressures}
\end{equation}

\begin{equation}
\vec{B}'=\left(1-\frac{m_e}{m_i}\nabla^2\right)\vec{B}\ ,
\vec{V}_e=\vec{V}_i-\vec{J}/n\ , \vec{J}=\nabla\times\vec{B}
\label{misc}
\end{equation}

Here $F_y=F_y(x,y,z,t)$ is the forcing function, the form of which is
discussed in below. For simplicity we assume an isothermal equation of
state for both electrons and ions (qualitatively similar to the
adiabatic case), and thus take $T_e$ and $T_i$ to be constant. The
normalizations of Eqns.~(\ref{mom})-(\ref{misc}) are based on constant
reference values of the density $n_0$ and the reconnecting component
of the magnetic field $B_{x0}$, and are given by (normalized $\to$
physical units): $t\to\omega_{ci}t$, $\omega_{ci}=eB_{x0}/(m_ic)$,
$x\to x/d_i$, $d_{i,e}=c/\omega_{pi,e}$, $\omega_{pi,e}^2=4\pi
n_0e^2/m_{i,e}$, $n\to n/n_0$, $\vec{B}\to\vec{B}/B_{x0}$,
$\vec{V}_{i,e}\to \vec{V}_{i,e}/V_{Ax}$,
$V_{Ax}=\omega_{ci}d_i=B_{x0}/(4\pi n_0 m_i)^{1/2}$, $T_{i,e}\to
T_{i,e}4\pi n_0/B_{x0}^2$, $p_{i,e}\to p_{i,e}4\pi/B_{x0}^2$,
$\vec{J}\to\vec{J}/(n_0eV_{Ax})$, and $F_y\to F_y d_i /(m_i n_0
V_{Ax}^2)$. Our algorithm employs fourth-order spatial finite
differencing and the time stepping scheme is a second-order accurate
trapezoidal leapfrog \cite{zalesak1, zalesak2} . We consider a cubical
3D simulation box with physical dimensions $L\times L \times L= 25.6
d_i \times 25.6 d_i \times 25.6 d_i$ (so that $L=25.6$ in normalized
units) and periodic boundary conditions imposed at $x= \pm L/2$,$y=
\pm L/2$, and $z= \pm L/2$ .  The simulation grid is $n_x \times n_y
\times n_z= 128 \times 256, \times 128$, yielding grid scales of
$\Delta_x=\Delta_z=0.2$ and $\Delta_y=0.1$.  The electron to ion mass
ratio is $m_e/m_i=1/25$ so that the normalized electron skin depth is
$d_e=\sqrt{m_e/m_i}=0.2$. The frozen-in law for electrons is broken primarily by
the presence of finite electron inertia, which is manifested by the
$m_e$ terms in the definition of $B'$ in Eq.~(\ref{misc}).  Very near
the x-points, however, the electron inertia terms become weak in
quasi-steady conditions, since both the $\partial_t$ and
$\vec{v}\cdot\nabla$ terms on the left-hand-side of Eq.~(\ref{mom})
become small. In these small regions the frozen-in law is mainly
broken by numerical diffusion. Past studies of this effect (\textit{e.g.} Ref.~\cite{Rem}) have found that the rates of reconnection are relatively insensitive to such grid-level
diffusion and are in reasonable agreement with the rates obtained from
particle simulations \cite{Uzdensky00}.

\subsection{Initial Equilibrium}
We begin with a one dimensional equilibrium. Three dimensional effects
will enter via the forcing function described in the next section. The
normalized magnetic field and density profiles in our initial
equilibrium are given by:
\begin{eqnarray}
\vec{B}(y) & = & \sin \left[\frac{2\pi}{L} \left( y + \frac{L}{4}\right)
\right]\vec{\hat{x}}  \\ 
n(y) & = & 1
+ \frac{1-B_x^2}{2(T_i+T_e)}
\end{eqnarray}
The boundary conditions are periodic in all three directions. As
required by these boundary conditions, $B$ is periodic under $y \to
y+L$.
Note that the normalization parameter $B_{x0}$ has been chosen so that
the peak value of this initial field is unity. From the given form of
$n$ it is apparent that $n=1$ at this location, or in physical units,
$n=n_0$.  The density profile is chosen to satisfy the total pressure
balance condition, which in normalized form is given by
\begin{equation}
n(T_i+T_e)+\frac{1}{2}B_x^2 ={\rm constant}
\end{equation}
Unless otherwise stated, we take the (constant) total temperature to
be $T_{tot}=T_i+T_e=1.0$, or in physical units $4\pi
n_0T_{tot}/B_{x0}^2=1$, so that the plasma $\beta$ based on the
reconnecting field has a (minimum) value of 2 where $B_x=1$ and
$n=1$. The density reaches a maximum value of $n=1.5$ in the center of
the equilibrium current sheets ($y=\pm L/4$) where $B_x=0$. Initially
the current is carried entirely by the electrons, while the ions begin
at rest.  To prevent energy buildup at the grid scale, the simulations
include fourth order dissipation in the density and momentum equations
of the form $\mu_4\nabla^4$ where $\mu_4=5.1\cdot10^{-5}$.  To avoid
physically artificial effects that can arise from exact reflection
symmetries of the initial condition, a small amount of random noise is
added to the magnetic field and ion current at the levels
$|\tilde{B}_{max}|\approx 10^{-4}$, $|\tilde{J}_{i_{max}}|\approx
10^{-4}$.

The linear tearing mode stability parameter is defined as $\Delta' =
[\partial_y \psit(y \to 0+) - \partial_y \psit(y \to 0-)]/\psit(y=0)$
where $\psit$ is the perturbation in flux the function \cite{FKR}.  In
our system it is given by \cite{Ott}:
\begin{eqnarray}
\Delta' (k_x) =- \left(2 k_0 \sqrt{\frac{k_x^2}{k_0^2}-1} \right) 
\tan \left(\sqrt{\frac{k_x^2}{k_0^2}-1}\right),
\end{eqnarray}
where $k_0=2\pi/L$. Since periodicity along the $x$ direction requires
$k_x \geq 2 \pi / L$ and thus $k_x^2/k_0^2 \geq 1$, one sees that
$\Delta' \leq 0$. Therefore, as noted in the introduction, the system
we consider is stable to tearing modes in the absence of forcing and
exhibits no spontaneous reconnection.

\subsection{Forcing Function}
\label{ForcingFunction}
The initial 1D equilibrium described in the previous section
does not contain a pre-seeded x-line. Rather, we employ a forcing
function to drive plasma and magnetic flux into the region surrounding
the point $(x,y,z)=(L/4,-L/4, 0)$, thereby forming an x-line along z
through that location. The forcing function has the general form
$F_y(x,y,t)= X(x)Y(y)Z(z)\Theta(t)$. Along x and z the forcing
function is a gaussian with widths $w_x$ and $w_z$, respectively,
centered at $(x,z) =(+L/4,0)$:
\begin{equation}
\begin{split}
X(x)&=\exp\left\{-\left[\frac{(x-L/4)}{w_x/2}\right]^2\right\}, \\ Z(z)&=\exp\left\{-\left(\frac{z}{w_z/2}\right)^2\right\}
\end{split}
\end{equation}
Along the inflow ($y$) direction, the forcing function is
antisymmetric in $y$ about $y=-L/4$.  It varies linearly with $y$
close to the reconnection point and then levels off to a constant
value ($Y\to \pm 1$) over a distance of $\sim 2 w_y$ upstream of the
x-point:
\begin{equation}
Y(y)=\tanh \left(\frac{y-L/4}{w_y} \right)-\tanh \left(\frac{y+L/4}{w_y} \right)+1
\end{equation}

The temporal behavior of the forcing is controlled by the function
$\Theta(t)$. This function starts at zero and increases monotonically
with time at the rate $d\Theta/dt\sim 1/\tau_f=0.1$ until it plateaus
at the value $F_\infty$:
\begin{equation}
\Theta(t)=F_\infty \tanh\left(\frac{t}{|F_{\infty}| \tau_f}\right)
\end{equation}
Since the initial slope of $\Theta(t)$ and hence the initial rate of
ramping is held constant at $1/\tau_f=0.1$ from one simulation to the
next, it takes more time to ramp up to the stronger levels of forcing
(roughly $\tau \sim |F_{\infty}|\tau_f$).

In our simulations the spatial and temporal structure of the forcing
is controlled by the parameters $w_x, w_y, w_z, \tau_f$, and
$F_\infty$. The appropriate choice of these parameters presumably
varies from system to system, depending on the physical application
under consideration and the nature of the forcing agent. In this
study, as a first step, we examine the reconnection behavior as a
function of the asymptotic strength of the forcing function
$F_\infty$, holding the other forcing function parameters fixed at
values that are convenient for the size of our simulations, $e.g.$
$w_x= w_z= 5.0 d_i, \tau_f=1.25\omega_{ci}^{-1}$ .  In the
substantially larger system considered in Ref. \cite{Sullivan},
the authors employed a proportionally broader forcing function with results
similar to those described here. The detailed study of how the geometry of 
the forcing function impacts the reconnection is left for future work.

\section{Numerical Results}
\label{results}
This study includes ten simulations. Five are two dimensional and
differ from each other only in the asymptotic level of forcing
$F_{\infty}$. These five simulations are included as a baseline for
comparison with the other five, which are three dimensional with
forcing localized as a gaussian in the out of plane direction. In the
3D simulations, the width of the forcing function along $z$ is equal to
its width along $x$ ($w_x=w_z=5.0d_i$). In the next section we will
describe the results of a typical 2D run and then compare to a similar
3D case. First, however, some important definitions are required.

Following Ref.~\cite{Shay04} and our previous 2D work~\cite{Sullivan},
we focus on the ion inflow $V_{in}\ (=V_y)$, outflow $V_{out}\
(=V_x)$, upstream reconnecting magnetic field $B_d\ (=B_x)$, and
upstream density $n_d$, near the boundaries of the ion dissipation
region where the ions decouple from both the electrons and the
magnetic field.  As found in earlier work~\cite{Sullivan},
the ion and electron velocities typically decouple approximately $0.5
c/\omega_{pi} $ upstream of the x-point below the center of the
forcing function, and we therefore measure the upstream quantities
$V_{in}$, $B_d$, and $n_d$ at this location $(+L/4,-L/4\pm
0.5,0)$. The reconnecting component of the magnetic field does not
vary strongly as a function of $y$ between 0.5 and 1.0 ion skin depths
upstream of the x-point; therefore, the results presented here are
not strongly dependent upon the method used to determine the upstream
edge of the dissipation region. This holds in both 2D and 3D.

The electron outflow ($i.e.$ outflow in the x-direction) typically
becomes quite large very close to the x-point and can be in excess of
the ion Alfv\'en speed. However, at the downstream edge of the
dissipation region, the electrons slow down to flow roughly with the
ions. The location where the two flows come together is typically
close to the location of the maximum in $V_{ix}$. Therefore the
downstream edge of the dissipation region is taken to be the location
of the maximum in $V_{ix}$.  Defining the downstream edge as the
location of the ion-electron velocity crossover point yields nearly
identical results \cite{Shay04, Sullivan}.

\begin{figure*}
\centering
\includegraphics[width=\textwidth]{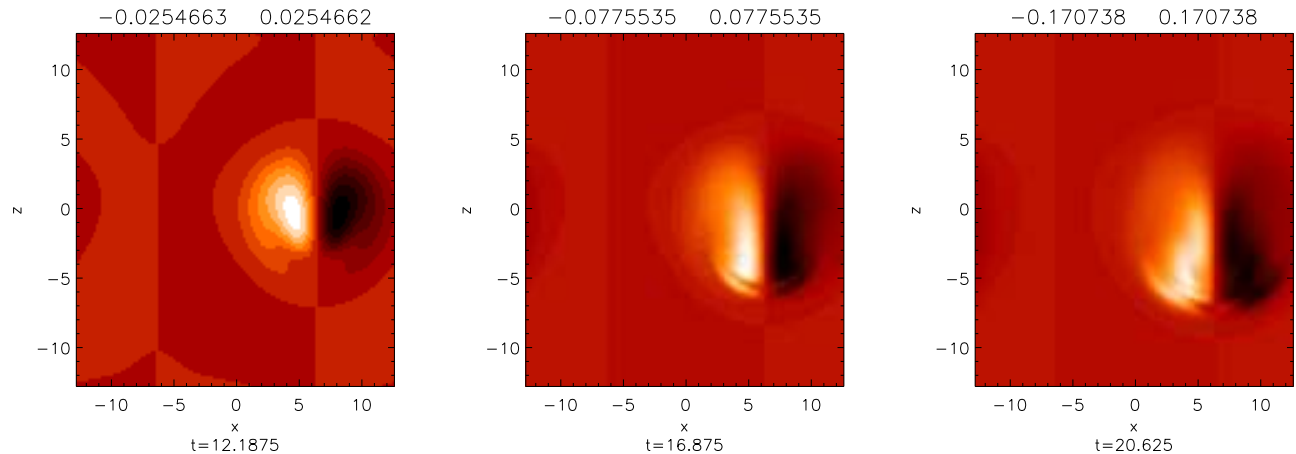}
\caption{(color online) Time evolution of $B_y(x,z)$ at $t=12.1875$ (left), $t=16.875$ (center), and $t=20.625$ (right). The reconnected flux moves in the $-z$ direction. Maximum and minimum values of the plotted quantity are shown above each plot. }
\label{by_multi}
\end{figure*}

A last important definition concerns the reconnection rate. In 2D,
the reconnection rate is defined as the time derivative of the
reconnected magnetic flux inside the magnetic island {\it per unit
length} in the (uniform) $z$ direction:
\begin{equation}
R_0 = \frac{\partial}{\partial t}\int_{x_o}^{x_x} B_y(x,-L_y/4) dx ,
\label{2dreconnectionrate}
\end{equation} 
where $y=-L/4$ is the plane where $B_x=0$, and $x_{_X}$, $x_{_O}$ are
the x-locations of the x-point and o-point, respectively.  A {\it
total} reconnection rate (rather than a rate per unit length) cannot
be defined in the 2D limit, since the uniformity of the system along
$z$ implies that the total reconnected flux is infinite. This is not
the case in the 3D simulations discussed here, in which the extent of the magnetic
island and the total reconnected flux are both finite in the $z$
direction. Thus in 3D it is natural to define the reconnection rate as
simply the time derivative of the {\it total} reconnected flux inside
the magnetic island:
\begin{equation}
R_{tot}=\frac{\partial}{\partial t} \int_A B_y dx\ dz
\label{3dreconnectionrate}
\end{equation}
where $A$ is the area of the magnetic island (found numerically using
field-line tracing) in the central plane of the equilibrium current
sheet $y=-L_y/4$. Fig.~(\ref{by_multi}) shows a time series $B_y$ in
this plane in a simulation discussed in the next section. The location
of the forcing function, centered at $(x,z)=(6.4,0)$, can be discerned
from the dipolar signature in the right half of the figures. As we
show below, nearly all the reconnected flux is inside this roughly
circular region under the forcing function.  (In both 2D and 3D,
small secondary islands are also typically present; however, the total
flux in these islands is small and they have no significant impact on
the reconnection rates reported here.)  The 3D reconnection rate given
by Eq.~(\ref{3dreconnectionrate}) has physical units of [flux/time],
in contrast to Eq.~(\ref{2dreconnectionrate}), which has physical
units of [flux/time/length]. To compare the 2D and 3D reconnection
rates, the 3D rate therefore must be somehow translated into a
reconnection rate per unit length, as in 2D. No unique way exists to
do this, making a meaningful comparison of the 2D and 3D reconnection
rates difficult. Two crude approaches to such a comparison, however,
are discussed further in the following section.
\subsection{Temporal Behavior}
\begin{figure*}[ht]
\centering
\includegraphics[width=\textwidth]{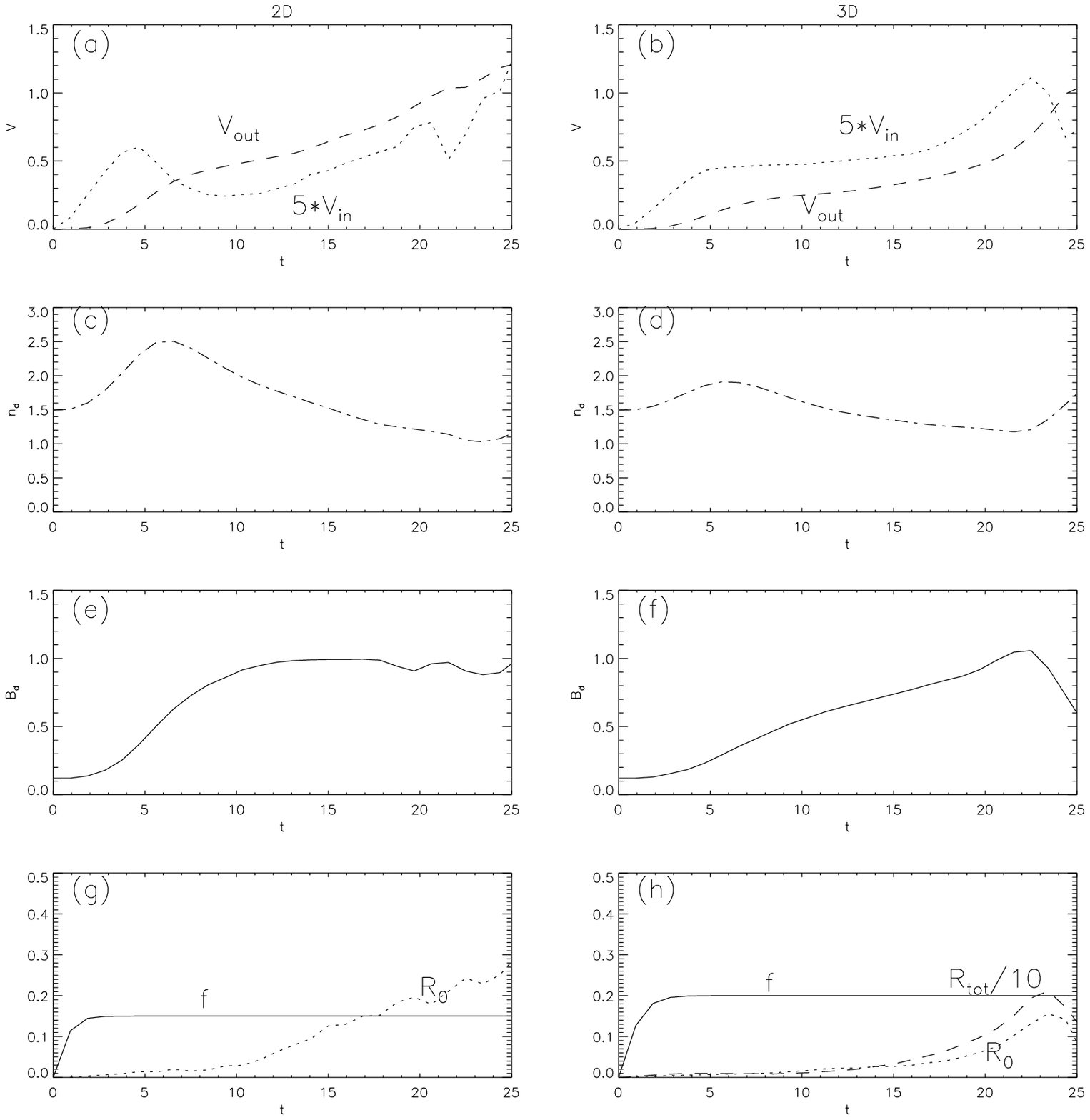}
\caption{ Time series of the dissipation region parameters in 2D(left) and 3D(right) with $F_\infty=0.15$, and $0.2$, respectively. }
\label{timeseries_2_15}
\end{figure*}

\begin{figure}[ht]
\centering
\includegraphics[width=0.5\textwidth]{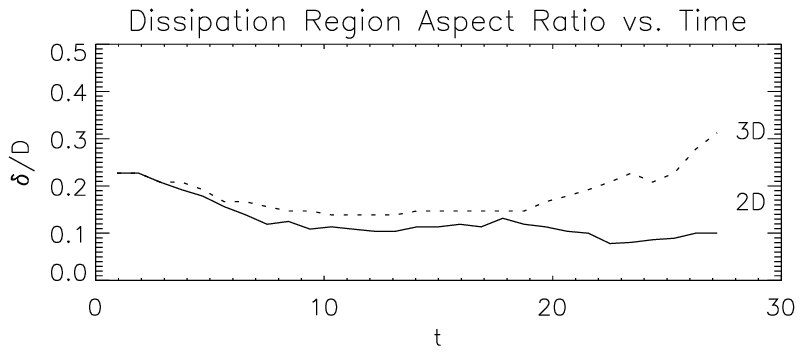}
\caption{Time series of the aspect ratio of the dissipation region vs. time in 2D (solid curve) and 3D (dashed curve. (Same simulations as the data shown in Fig.(~\ref{timeseries_2_15}))}
\label{aspect}
\end{figure}

Fig.~(\ref{timeseries_2_15}) shows time series of the dissipation
region parameters in typical 2D (left) and 3D (right) simulations. In
the 2D simulation, the asymptotic upstream forcing strength is
$F_\infty = 0.15$, as seen from the solid curve in
Fig.~(\ref{timeseries_2_15}g). Note the level of forcing has become
constant by $t \approx 3$. In panel (a), we plot the outflow velocity
(dashed curve), and 5 times the inflow velocity (dotted curve) in the
2D case (the factor of 5 serves to place the quantities on a similar
scale). As the forcing is turned on, the inflow velocity increases,
causing the pressure and density (panel c) to mount under the forcing
function. This back pressure causes the inflow velocity to level off
and then decrease around $t \approx 4.5$. Magnetic flux also piles up
under the forcing function causing the upstream field to increase, as
seen in panel (e) during the period $t<12$.  Consistent with past
studies ($e.g.$ Refs. ~\cite{Shay04, Sullivan, Cassak}), at about $t\sim 12$,
when the flux pile-up has narrowed the current sheet to about an ion
skin depth, the layer opens up into a Petschek-like configuration and
fast reconnection begins, as can be seen by the rise in the
reconnection rate $R_0$ in panel (g).  This onset of reconnection
causes $V_{in}$ to increase again and $B_d$ to level off, indicating
that the rate of reconnection is sufficient to prevent further pile up
of upstream flux, $i.e.$ flux is liberated from the dissipation region
by reconnection at a rate comparable to that with which flux is pumped
into the region, so that $B_d$ is stable for an extended period ($t
\sim 10-25$). The ratio $V_{in}/V_{out}\sim 0.1-0.2$ in this phase is
consistent with continuity arguments: Assuming in the 2D case that the
ion diffusion region has a width $2\delta$ and a length $2D$, the
approximate incompressibility of the ion flow yields $V_{in}D\sim
V_{out}\delta$ or $V_{in}/V_{out}\sim \delta/D$.  As shown in
Fig.~(\ref{aspect}), the value of $\delta/D$ is typically between 0.1
- 0.2. Here $\delta/D$ is computed by dividing the half-width at
half-max of the current sheet at the location of the x-line by the
distance from the x-line to the location of the peak ion outflow
velocity along $x$.  The aspect ratio varies somewhat in time,
approaching some minimum value then generally becoming slightly larger
again as the current sheet opens up and shortens into a Petschek-like
configuration.
Strong reconnection continues in the 2D system for about another $10$
time units beyond the final time ($t=25$) shown in the figures, at
which point the system runs out of magnetic flux to reconnect and the
reconnection rate drops sharply to zero.  The features described here
are observed at all levels of forcing in 2D that are sufficiently
strong to trigger Alfv\'enic reconnection. In Ref.
\cite{Sullivan} qualitatively similar results were found in a larger
($L \times L = 102.4 d_i \times 102.4 d_i$) 2D system.
Several differences are apparent in comparing the 2D data shown in left half of
Fig.~(\ref{timeseries_2_15}) with the 3D data in the right half. For example, the enhancement in the density
(panel d) and upstream magnetic field (panel f) are weaker or more
gradual in 3D than in 2D. This results from a key difference between
2D and 3D: in 2D flux can compress upstream of the current sheet or it
can reconnect, flowing out along $x$. Similarly, density can build up
under the forcing function or it can be cleared out by
reconnection. In 3D, plasma and magnetic flux have the option of
either reconnecting or flowing away from the forced zone along the
$\pm z$ directions. This difference has at least two important
consequences: first, the forcing function is not as effective at
compressing plasma and flux in 3D. Indeed, the gradual rise in $B_d$
seen in Fig.~(\ref{timeseries_2_15}f) results largely from the
convection of stronger upstream magnetic field into the reconnection
zone rather than from in-situ compression. To partially compensate for
this, in Fig.~(\ref{timeseries_2_15}) we compare 2D and 3D simulations
with somewhat different levels of forcing ($F_\infty=0.2$ in 3D
vs. $F_\infty=0.15$ in 2D as shown in
Figs.~(\ref{timeseries_2_15}g,h)). A second consequence, which we
discuss further below, is that the reconnection ``efficiency'' in 3D
is smaller than in 2D -- that is, a smaller fraction of the upstream
magnetic flux that is initially contained within the forced region is
reconnected in 3D than in 2D, making the reconnecting phase of the 3D
simulations shorter than those in the 2D runs.

\begin{figure}
\includegraphics[width=0.5\textwidth]{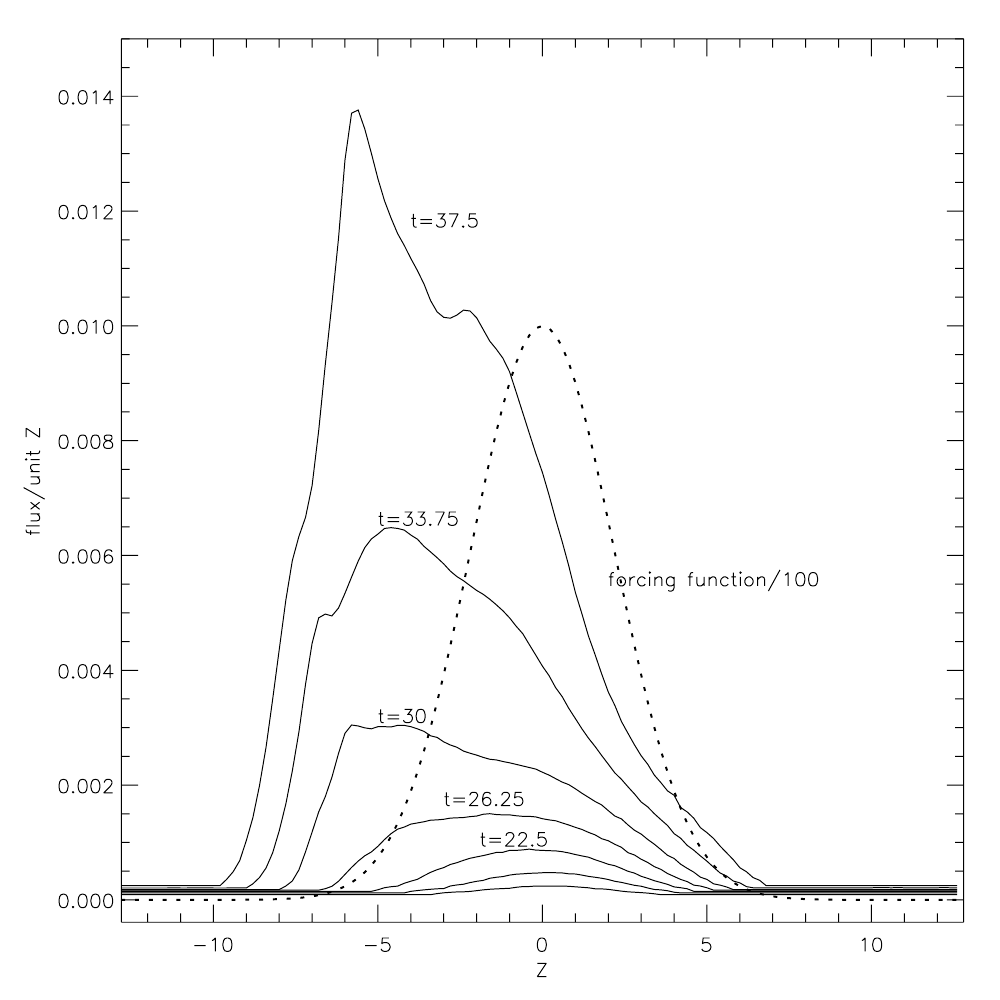}
\caption{Time evolution of the reconnected flux per unit z as a function of z (solid curves), and the shape of the forcing function(dashed curve).}
\label{fluxevolution}
\end{figure}

Two measures of the reconnection rate are shown in
Fig.~(\ref{timeseries_2_15}h). The dotted curve labeled $R_0$, like
panel (g), is the integral given by Eq.~(\ref{2dreconnectionrate})
calculated in the $z=0$ plane. This quantity is a rough measure of the
rate of reconnection immediately under the center of the forcing
function. Although one might expect this $z=0$ location to yield the
largest local reconnection rate, in fact substantially larger values
can occur at negative $z$ values near the edge of the forced
zone. This can be seen from Fig.~(\ref{fluxevolution}), which shows
the $z$-dependence of the reconnected flux per unit $z$
[Eq.~(\ref{2dreconnectionrate}) without the time derivative] at
various times. The area under the curves at a given time yields the
total reconnected flux [Eq.~(\ref{3dreconnectionrate}) without the
time derivative].  The flux distributions become skewed to the left at
later times due to the dragging of the magnetic field by the
electrons. This leftward electron flow carries the bulk of the
(rightward) electric current density that supports the reversal in the
reconnecting magnetic field $J_z\propto dB_x/dy$.  The transport of
reconnected flux by the electron flow can also be seen in the downward drift of the $B_y$ profile in the center of the sheet shown in Fig.~(\ref{by_multi}). This effect has also been reported in other studies of three-dimensional reconnection (see, for example, Refs.~\cite{Rudakov02, Huba02, Hesse05, Lapenta06}).  
A second measure of the
reconnection rate is shown as the dashed curve in
Fig.~(\ref{timeseries_2_15}h) labeled $R_{tot}/10$. This curve depicts
the total reconnection rate $R_{tot}$ given by
Eq.~(\ref{3dreconnectionrate}) divided by the full width of the
forcing function along $z$ ($2w_z=10.0d_i$). As can be seen from
Fig.~(\ref{fluxevolution}), this width roughly characterizes the
extent along $z$ of the reconnected flux distribution and thus
$R_{tot}/10$ is a crude measure of the average reconnection rate per
unit length in the 3D system.  Although the reconnection rates as
characterized by these measures are somewhat smaller in the 3D system
than those in the 2D case, we show in the the following section that,
when the 2D and 3D reconnection rates are plotted as a function of the
instantaneous values of the upstream reconnecting magnetic field
[Figs.~(\ref{timeseries_2_15}e,f)], similar results are obtained.
That is, for a given value of the upstream magnetic field, the rates
of reconnection characterized by either $R_0$ or $R_{tot}/10$ are
roughly comparable to the corresponding 2D values.
Another feature apparent in the 3D data of
Fig.~(\ref{timeseries_2_15}) is the downturn in the reconnection rate
(panel h) and upstream field (panel f) seen at $t\simeq 23$. At this
time, nearly all of the magnetic flux in the forced zone has been
expelled by the forcing function, and thus the system has effectively
run out of flux to reconnect.  This is much earlier than the otherwise
similar collapse in the 2D reconnection rate at $t\simeq 38$. As was
noted earlier, less flux is reconnected in 3D because the plasma and
magnetic field are free to flow outward along the $\pm z$ directions
without reconnecting. In 2D, the uniformity of the system along $z$
prevents any net transport of flux in that direction, and a much
larger fraction of the initial upstream flux in 2D (over 70\% percent
compared to about 10\% in 3D) ends up inside the magnetic
islands at late times.

\begin{figure}[h]
\begin{center}
\includegraphics[width=0.5\textwidth]{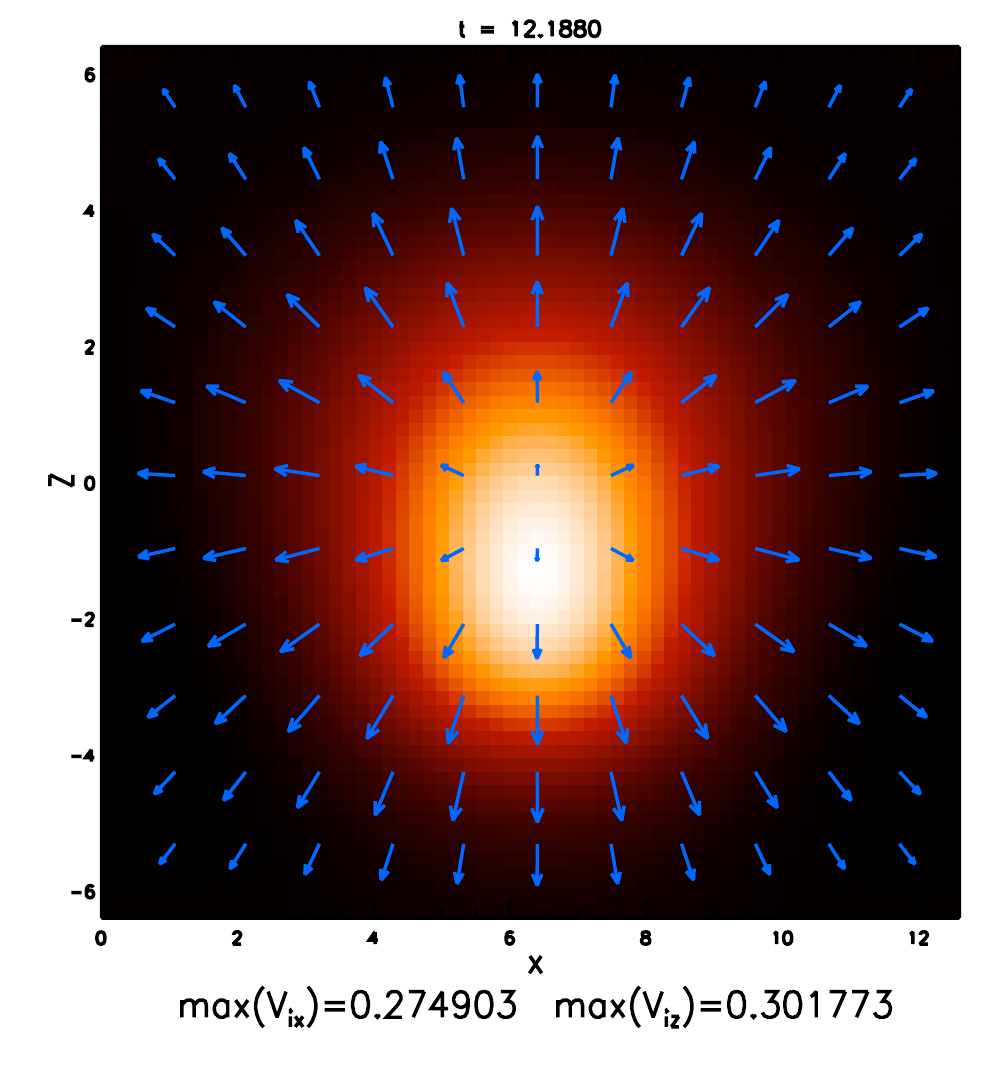}
\caption{(color online) The gray scale plot in this figure shows $J_z$ in the central plane of the forced current sheet at $t=12.188$ in the 3D simulation with $F_\infty=0.2$. The over-plotted
vectors (blue online) show the relative magnitude and direction of $\vec{V}_i=\vec{J}_i/n$ in the same plane. }
\label{ion_outflow}
\end{center}
\end{figure}

A final difference between 2D and 3D evident
from Figs.(\ref{timeseries_2_15}a,b) is that the ratio
$V_{in}/V_{out}$ in 3D is enhanced by roughly a factor of two over the 2D
case.  Given that the layer aspect ratio $\delta/D$
[Fig.~(\ref{aspect})] shows a much smaller increase, this enhancement
is inconsistent with the 2D continuity argument described earlier,
$V_{in}/V_{out}\sim \delta/D$. In the 3D simulations, however, the
continuity relation must be modified by a geometric factor.  In 2D,
the ion outflow from the reconnection region is almost entirely along
the $x$ direction with $V_{ix}\gg V_{iz}$. In 3D, as one can
see from the snapshot shown in Fig.~(\ref{ion_outflow}), the ion
outflow is nearly omnidirectional with $V_{iz}\sim V_{ix}$, and the
ion diffusion region is disk-shaped (like a hockey puck) rather than
rectangular.  Assuming this disk has radius $D$ and half-thickness
$\delta$, plasma will thus flow into an area $A_{in}=\pi D^2$ (one
surface of the disk) and out through an area $A_{out}= 2 \pi D
\delta$. Continuity then demands that $V_{in}A_{in}=V_{out}A_{out}$ or
\begin{equation}
V_{in}= \left(\frac{A_{out}}{A_{in}}\right)V_{out}
=\left(\frac{ 2 \pi D \delta}{\pi D^2}\right)V_{out}
=\left( \frac{2\delta}{D} \right)V_{out}.
\label{3dcontinuity}
\end{equation}
Thus, for the same aspect ratio $\delta/D$, one would expect
$V_{in}/V_{out}$ to be a factor of two larger in the 3D case, as 
indeed it is.
\subsection{Scaling of Reconnection}
\begin{figure*}[ht]
\centering
\includegraphics[width=1\textwidth]{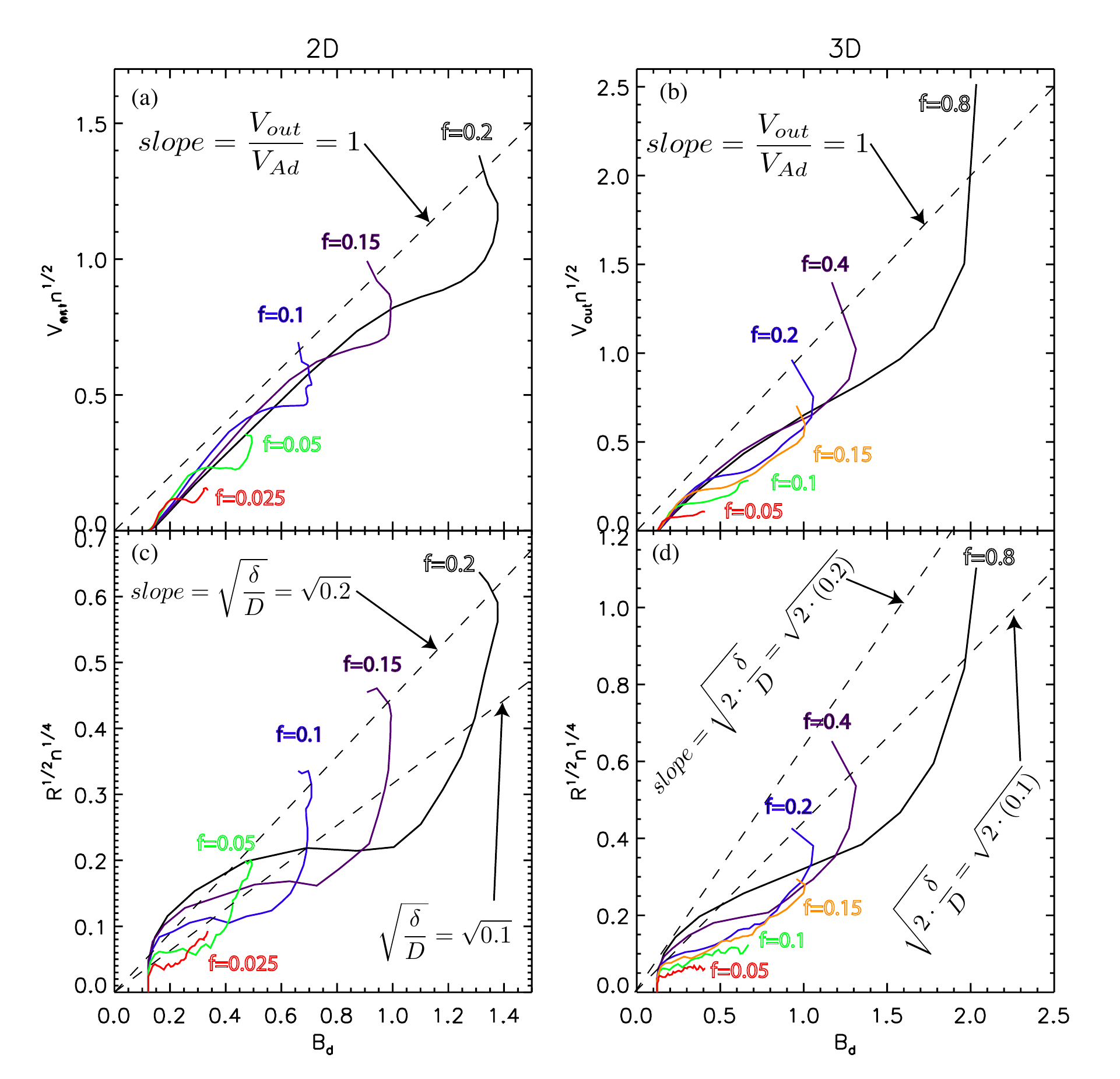}
\caption{(color online) (a) $V_{out} \sqrt{n} $ vs. $B_d$ in 2D, (b) $V_{out} \sqrt{n}$ vs. $B_d$ in 3D , 
(c) $\sqrt{R_0}\sqrt[4]{n} $ vs. $B_d$ in 2D, (d)$\sqrt{R_0} \sqrt[4]{n} $ vs. $B_d$ in 3D  }
\label{Scaling3Dand2D}
\end{figure*}

In this section we test two simple scaling relations for the plasma
outflow and reconnection rate that have been found in past
studies to characterize reconnection in some forced and unforced 2D systems
($e.g.$ \cite{Shay99,Shay04,Huba04,Shay98b,Hesse99,Birn01,Shay01,Sullivan}). In
the case of the outflow, the dynamics that cause reconnected field
lines to accelerate plasma away from the x-point are similar to those
of an Alfv\'en wave, and so $V_{out} \sim V_{Ad}$ where
$V_{Ad} = B_d/ \sqrt{4 \pi m_i n}$. Regarding the reconnection rate,
for a given dissipation region of length $D$ and thickness $\delta$,
the continuity arguments discussed in the last section yield $V_{in}
\sim C_0 (\delta/D)V_{Ad}$ where $C_0\sim 1$ in 2D and $C_0\sim 2$ in
3D. Assuming this inflow along $y$ carries an $x$-directed magnetic
field $B_d$ into the dissipation region, the rate of reconnection per
unit length along $z$ (equivalent to the rate at which magnetic flux
enters the reconnection zone per unit length along $z$) is $V_{in}
B_d$. We therefore expect:
\begin{subequations}
\begin{align}
V_{out}  &\sim \left(\frac{\delta}{D}\right) \frac{B_d}{ \sqrt{4 \pi m_i n}}   \\
R_0 &\sim V_{in} B_d \sim \left(\frac{C_0\delta}{D}\right) \frac{B_d^2}{\sqrt{ 4 \pi m_i n}}\ .
\end{align}
\end{subequations}
Note the reconnection rate $R_0$ is predicted to scale
with the \emph{square} of the upstream field.
In normalized units, isolating the upstream field $B_d$ as the
independent variable, these become
\begin{subequations}
\begin{align}
V_{out} n^{1/2} &\sim \left(\frac{\delta}{D}\right) B_d  \label{VoutScaling} \\
R_0^{1/2} n^{1/4} &\sim \left(\frac{C_0\delta}{D}\right)^{1/2} B_d \label{ratescaling}
\end{align}
\end{subequations}
We now test these scaling laws under the assumption that $\delta/D$
has an approximately constant value (on the order of 0.1 - 0.2) over
the duration of the Alfv\'enic phase of the reconnection process. In
Figs. ~(\ref{Scaling3Dand2D}a,b), to test the validity of
Eq.~(\ref{VoutScaling}), the quantity $V_{out} \sqrt{n_d}$ is plotted
versus $B_d$ with time as a parameter. Each curve represents data from
a single simulation and is labeled by the value of $F_{\infty}$ for
that simulation. A sixth 2D simulation with $F_{\infty}=0.025$ is
included to extend the range of data. (This weakest level of forcing
is insufficient to produce Alfv\'enic reconnection in 3D, so no 3D
analog for this simulation is included.)  The dotted line of slope
unity, plotted for reference, represents an outflow velocity equal to
the upstream Alfv\'en speed based on $B_d$. At the stronger forcing
levels, the data are indeed in rough agreement with the predicted
scaling, albeit with significant, order-unity time variations within
any given simulation.  At the weakest forcing levels, the outflow
(like the reconnection rate discussed in a moment) falls below the
predicted scaling.  The forcing strength in these simulations is
insufficient to narrow the current sheet down to the ion skin depth
scale, and fast reconnection is never triggered.
In Figs.~(\ref{Scaling3Dand2D}c,d), the quantity $R_0^{1/2}n_d^{1/4}$
is plotted vs. $B_d$ to test the scaling of the reconnection rate
given by Eq. (\ref{ratescaling}). Plots based on $R_{tot}/10$ rather
than $R_0$ yield slightly larger but similar results. 
The dotted lines represent the expected scaling, assuming
that $V_{out}/V_A=1.0$ and that the system has an unvarying
dissipation region aspect ratio of $(\delta/D) = 1/5$ (steeper line)
or $(\delta/D) = 1/10$ (less steep). At the later times when the
reconnection rates are highest, the data in the more strongly forced
simulations roughly follow the expected trend. In the most weakly
forced runs, however, the scaling progressively breaks down; the
perturbation of the current sheet in these cases is insufficient to
trigger fast reconnection before a nearly force-balanced state is
reached.

\section{Conclusions}
\label{conclusions}
We have examined the scaling behavior of reconnection in a forced,
three-dimensional, periodic system using two-fluid simulations with
finite electron inertia. The forcing in the simulations was driven by
a spatially localized forcing function added to the ion momentum
equation inside the computational domain. We initialized the system
with a one dimensional, broad current sheet equilibrium that was
tearing-mode stable; three dimensional structure entered through the
localization of the forcing function. Comparisons were made to
analogous two-dimensional simulations.

In the two dimensional case, as found our previous work in larger 2D
systems \cite{Sullivan}, sufficiently strong levels of forcing were
found to produce a quasi-steady Petschek-like reconnection
configuration with a dissipation region aspect ratio $\delta/D\sim
0.1-0.2$. The forcing function produces a pileup of magnetic flux in
the upstream region and hence an increase in the upstream magnetic
field, until the reconnection rate becomes sufficient to prevent
further pileup of magnetic flux. The flux pile-up typically halts once
a relatively thin current sheet is formed---between 0.5 and 1.0 ion
skin depths in width---at which point the magnetic separatrix opens
up, and fast, Alfv\'enic reconnection begins, leading to a period when
the upstream reconnecting field $B_d$ is relatively stable. The rate
of reconnection per unit length along the (ignorable) $z$ direction
was found to scale roughly like $B_d^2$, as expected from
simple scaling arguments, although significant time variations
are observed during the course of any given run than cannot
be explained by this simple scaling.

In three dimensions, freedom of the plasma to flow out in the $\pm z$
directions makes the forcing function less effective at compressing
plasma; the resulting flux pile-up is weaker in 3D, convection
dominates over compression in the upstream region, and the period of
reconnection is shorter than in comparable 2D systems. The
localization of the magnetic island along $z$ in 3D makes it possible
to define a total reconnection rate, rather than a rate per unit
length, as in 2D. For sufficiently strong forcing, this total rate,
like in 2D result, was found to be roughly proportional $B_d^2$.

To directly compare the 2D and 3D reconnection rates one must
translate the 3D rate into a rate per unit $z$, as in 2D. Two rough
methods of extracting such a quasi-2D rate from the 3D data were
described -- one of which is simply to divide the total rate in 3D by
the width of the forcing function -- with similar results. When either
of these two rates are plotted as a function of the upstream magnetic
field, the 2D and 3D results are comparable. On the other hand, the
distribution of reconnected magnetic flux in 3D becomes strongly
skewed at late times by the flow of electrons in the current layer,
and near the edge of the forcing function, the flux builds at local
rates that can substantially exceed the 2D values. 

\begin{acknowledgments}
This  work was supported by NSF grant 0238694, 
DOE grant DE-FG02-07ER54915, NASA grant NNX07AR49G, and EPSCoR. Simulations were done at Dartmouth College.
\end{acknowledgments} 

\renewcommand{\baselinestretch}{1.0}


\end{document}